\documentclass[twocolumn, amsmath, amssymb, showpacs, prx]{revtex4-1}

\usepackage{amsmath}
\usepackage{graphicx}
\usepackage{hyperref}
\usepackage[usenames,dvipsnames,svgnames,table]{xcolor}
\definecolor{light-gray}{gray}{0.8}

\begin{document}

\title{\Large \bf{Tie Strength Distribution in Scientific Collaboration Networks}}

\author{Qing Ke}
\affiliation{Center for Complex Networks and Systems Research, School of Informatics and Computing, Indiana University, Bloomington, IN, USA}
\author{Yong-Yeol Ahn}
\email{yyahn@indiana.edu}
\affiliation{Center for Complex Networks and Systems Research, School of Informatics and Computing, Indiana University, Bloomington, IN, USA}

\date{\today}

\begin{abstract}
Science is increasingly dominated by teams. Understanding patterns of
scientific collaboration and their impacts on the productivity and evolution of
disciplines is crucial to understand scientific processes. Electronic
bibliography offers a unique opportunity to map and investigate the nature of
scientific collaboration. Recent work have demonstrated a counter-intuitive
organizational pattern of scientific collaboration networks: densely
interconnected local clusters consist of weak ties, whereas strong ties play
the role of connecting different clusters. This pattern contrasts itself from
many other types of networks where strong ties form communities while weak ties
connect different communities. Although there are many models for collaboration
networks, no model reproduces this pattern. In this paper, we present an
evolution model of collaboration networks, which reproduces many properties of
real-world collaboration networks, including the organization of tie strengths,
skewed degree and weight distribution, high clustering and assortative mixing.
\end{abstract}

\maketitle

\section{Introduction} \label{sec:intro}

Teams are increasingly overshadowing solo authors in production of knowledge~\cite{Wuchty:sci07}.
Examining patterns of scientific collaboration is therefore crucial to understand the scientific
processes, knowledge production~\cite{Wuchty:sci07}, research productivity~\cite{Beaver:79}, the
evolution of disciplines~\cite{Sun:scirep13}, and scientific impact~\cite{Jones:sci08, Uzzi:sci13},
etc. Electronic bibliographic data and the development of network science make it possible to
systematically investigate scientific collaboration at a large scale~\cite{Newman:review, BA:review,
Boccaletti:PR06:cx}. A common approach to studying scientific collaboration is to construct a
network of collaboration, where nodes represent authors and two authors are connected by
co-authorship~\cite{Newman:PNAS01:ca}. Various aspects of collaboration networks have been widely
explored, including basic structural properties~\cite{Newman:PNAS01:ca, Barrat:PNAS},
evolution~\cite{Barabasi:PA02:sci-net}, robustness~\cite{Holme:PRE02, Liu:plosone11}, assortative
mixing~\cite{Newman:PRL02:assortative}, and rich-club ordering~\cite{Colizza:NPhy06:rich, Opsahl:PRL08}.
Since coauthors usually know each other, collaboration networks have often been considered as proxies
of social networks~\cite{Newman:PNAS01:ca}. This viewpoint has been widely adopted, because
collaboration networks can be systematically constructed without any subjective
bias~\cite{Newman:PNAS01:ca} and the size of these networks can be large.

However, recent studies have revealed that collaboration networks possess unique properties that are
not presented in other proxies of real-world social networks such as mobile communication networks and
online social networks. One example is the atypical distribution of weak and strong ties. 
Like most other networks, collaboration networks exhibit cohesive groups (`communities')~\cite{girvan:community, fortunato:community, rosvall:infomap, ahn:link}. 
Since Granovetter pioneered the ideas of the relationship between network structure and tie strength, it has been assumed that strong ties tend to exist in the communities, while weak ties tend to connect these groups~\cite{Granovetter, girvan:community}. 
Here we refer `communities' in a purely structural point of view, ignoring weights and `weak' and `strong' ties refer the weight of edges. 
This organizational principle has been repeatedly confirmed in many networks~\cite{Friedkin:SN80, Onnela:pnas07, Onnela:njp07, Cheng:jsm10, Grabowicz:pone12, Pajevic:nphy12}. 
However, scientific collaboration networks exhibit the opposite pattern; weak ties constitute communities, while strong ties connect these research communities~\cite{Pan:EPL12:ca, Pajevic:nphy12}. 
This counter-intuitive observation raises a question: How and why collaboration networks are shaped in this way?

Although there are many models of scientific collaboration networks or similar weighted networks~\cite{Newman:PRE03:social, Catanzaro:PRE04:assort, Ramasco:pre04:ca, Barabasi:PA02:sci-net, Borner:pnas04:ca, Guimera:science05:team, Sun:scirep13, goh:pre05:weighted, goh:pre06:bbs}, the organization of tie strength and their roles on global connectivity have not been fully explored.
Here we propose that the academic advising system, the patterns of academic career trajectory, and the active inter-group collaboration may provide an explanation.
Our key notion is that weak ties are mainly formed from short-term collaborations between students and their advisors, while strong ties are formed through long-term collaborations between groups~\cite{Pan:EPL12:ca}.
Built on this notion, our model reproduces the tie-strength distribution as well as other common properties, such as skewed degree and weight distribution, high clustering, and assortative mixing.

\section{Structure and Link Weight}

To test the universality of the atypical tie-strength distributions in scientific collaboration reported in~\cite{Pan:EPL12:ca, Pajevic:nphy12}, we analyze four scientific collaboration networks: Network Science, High-energy Physics, Astrophysics, and Condensed Matter. Link weights in these networks are defined by $w_{ij} = \sum_p \frac{1}{n_p-1}$, where $n_p$ is the number of authors in paper $p$ in which $i$ and $j$ participated~\cite{Newman:PRE01:ca, Barrat:PNAS}.  Although this particular definition of weight is not unique, it has been widely accepted as a standard metric (See Section III in~\cite{Newman:PRE01:ca} for a detail and thorough discussion about it). Table \ref{tab:datasets} lists basic statistics of these networks. As many studies demonstrated, both degree and link weights are broadly distributed~\cite{Newman:PNAS01:ca, Barrat:PNAS}.

\begin{table*}
\centering
\begin{tabular}{c||c|c|c|c|c|c||c}
\hline
Name     & $N$    & $M$     & $\left< k \right>$ & $\left< w \right>$ & $c$   & $r$     & Time       \\ \hline
\hline
Net-sci  & 379    & 914     & 4.823              & 0.536              & 0.741 & -0.0817 &  -         \\ \hline
Hep-th   & 5,835  & 13,815  & 4.74               & 0.990              & 0.506 & 0.185   & 1995 -- 1999 \\ \hline
Astro-ph & 14,845 & 119,652 & 16.12              & 0.279              & 0.670 & 0.228   & 1995 -- 1999 \\ \hline
Cond-mat & 36,458 & 171,735 & 9.42               & 0.506              & 0.657 & 0.177   & 1995 -- 2005 \\ \hline
\end{tabular}
\caption{Structural statistics for weighted scientific collaboration networks
include number of nodes $N$, number of links $M$, mean node degree
$\left< k \right>$, mean link weight $\left< w \right>$, clustering coefficient
$c$~\cite{Watts:nature98}, and assortativity coefficient $r$~\cite{Newman:PRL02:assortative}.
Net-sci is based on the coauthorship of scientists working on network
science~\cite{Newman:PRE06:comm}. Hep-th, Astro-ph,
and Cond-mat are constructed based on the papers posted on High-energy Physics
E-Print Archive (\url{http://arxiv.org/archive/hep-th}), Astrophysics E-Print
Archive (\url{http://arxiv.org/archive/astro-ph}), and Condensed Matter E-Print
Archive (\url{http://arxiv.org/archive/cond-mat}),
respectively~\cite{Newman:PNAS01:ca}. For each network, we only consider the largest
connected component. All the 4 networks are downloaded from
\url{http://www-personal.umich.edu/~mejn/netdata/}.}
\label{tab:datasets}
\end{table*}

\begin{figure}[t]
\begin{center}
\includegraphics[trim=0mm 5mm 0mm 0mm, width=\columnwidth]{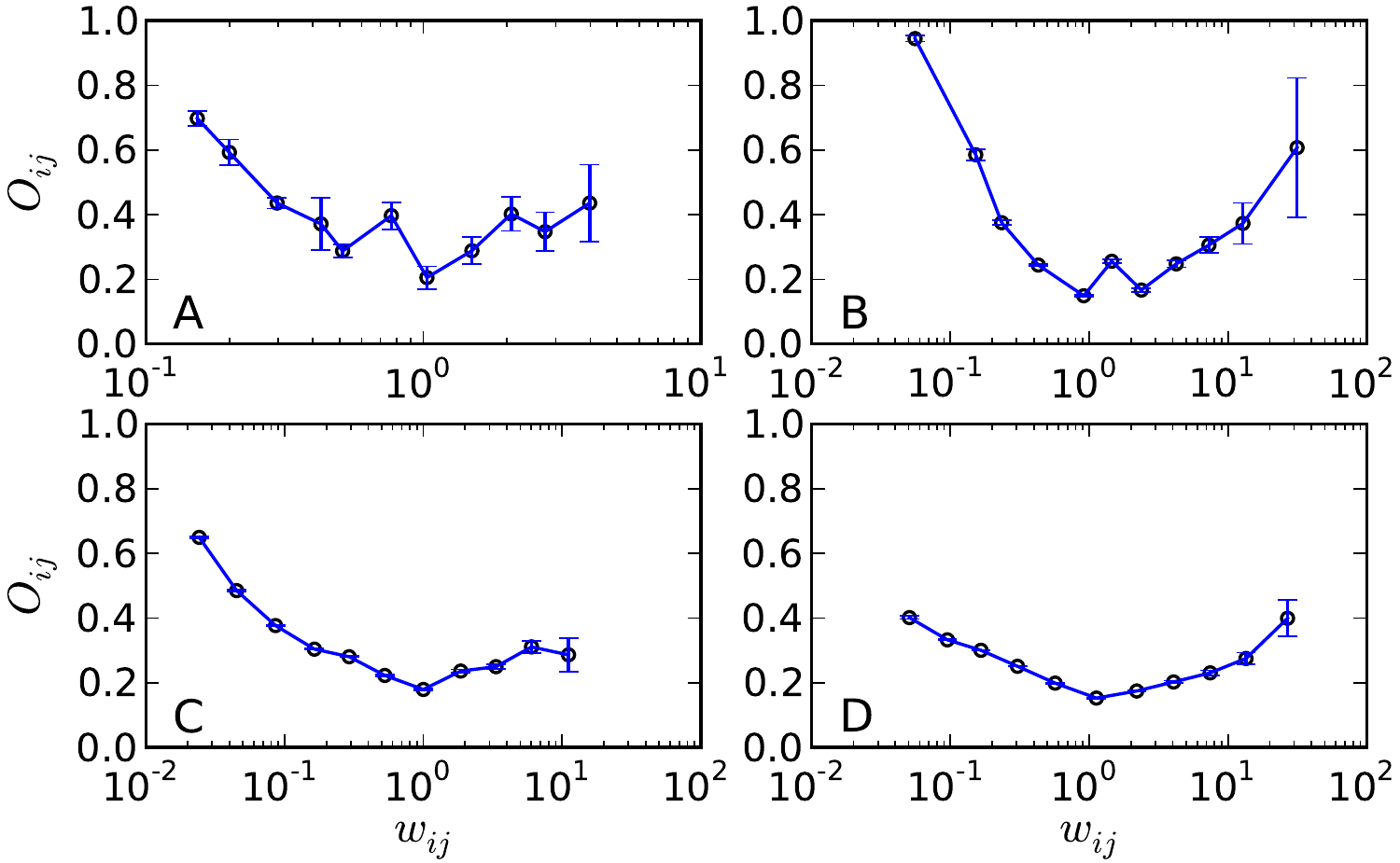}

\caption{\label{fig:wei_overlap}The correlation between link overlap $O_{ij}$
and link weight $w_{ij}$ in scientific collaboration network of (\emph{A}) Network
Science, (\emph{B}) High-energy Physics, (\emph{C}) Astrophysics, and (\emph{D})
Condensed Matter. We use logarithmic binning for $w_{ij}$. The error bars indicate
the standard error of the mean $O_{ij}$. For a large portion of links, overlap
decreases with weight. For a small portion of strongest links, overlap increases
with weight.}

\end{center}
\end{figure}

\begin{figure}[t]
\begin{center}
\includegraphics[trim=0mm 5mm 0mm 0mm, width=\columnwidth]{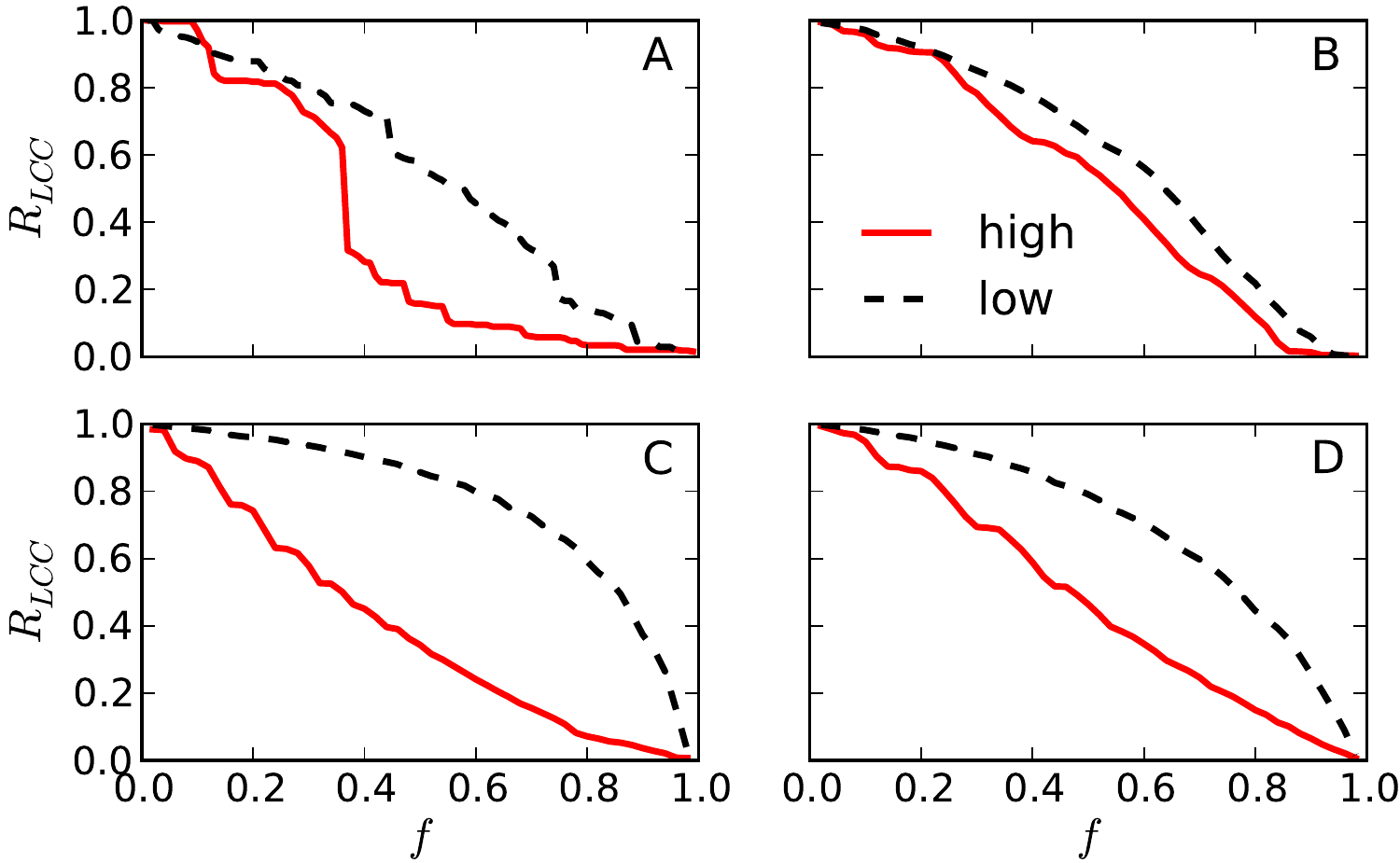}

\caption{\label{fig:link_rmv}The robustness of scientific collaboration network 
of (\emph{A}) Network Science, (\emph{B}) High-energy Physics, (\emph{C}) 
Astrophysics, and (\emph{D}) Condensed Matter under the removal of strong (weak) 
ties. The control parameter $f$ means the fraction of removed links. The removal 
of links is on the basis of their strength $w_{ij}$. The black dashed curves 
correspond to the removal of links from weak to strong. The red solid curves 
correspond to the removal of links from strong to weak. The relative size of 
Largest Connected Component (LCC) $R_{LCC} = N_{LCC}/N$ indicates that the 
removal of strong links leads to a faster breakdown of networks.}

\end{center}
\end{figure}

\begin{figure}[t]
\begin{center}
\includegraphics[trim=0mm 5mm 0mm 0mm, width=\columnwidth]{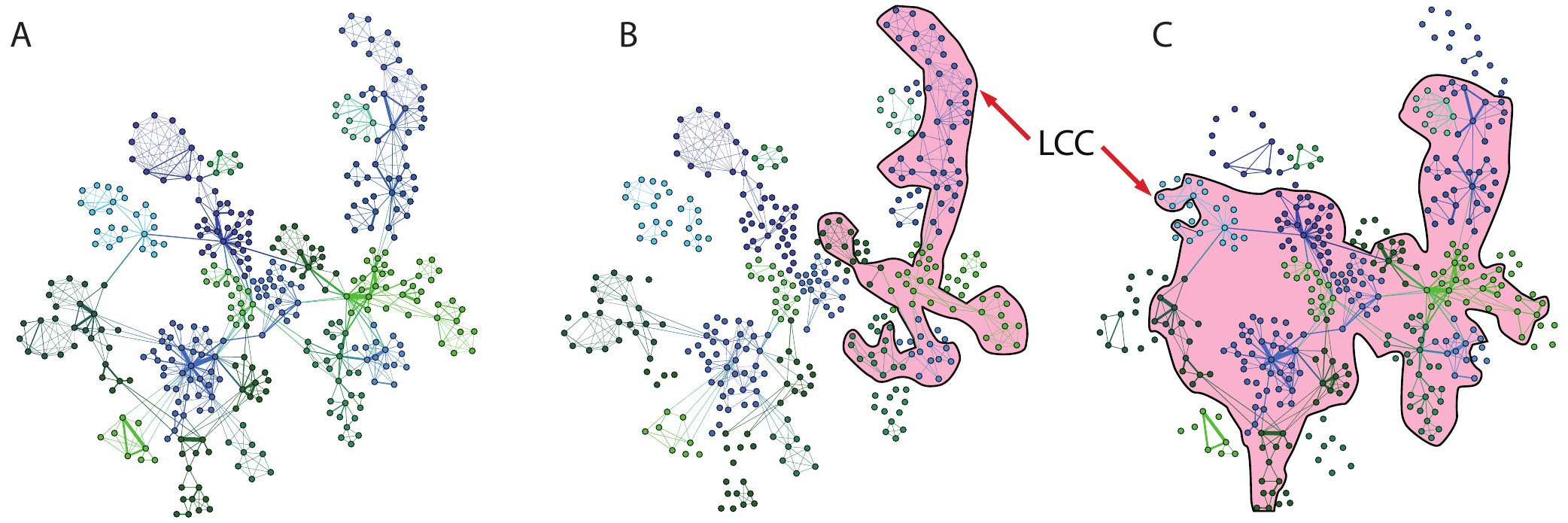}

\caption{\label{fig:net-sci}Visualization of the structure of Network Science
collaboration network and link removal process. (\emph{A}) The whole network
structure with 379 nodes and 914 links. The color of each node indicates its
community membership obtained by Louvain method~\cite{Blondel:Louvain}.
(\emph{B}) The remaining subgraph after removal of $43\%$ strongest links.
The shaded region indicates Largest Connected Component. (\emph{C}) The 
remaining subgraph after removal of $43\%$ weakest links.}

\end{center}
\end{figure}

Figure~\ref{fig:wei_overlap} shows the relationship between link weight
$w_{ij}$ and local clustering defined by the overlap measure $O_{ij} =
\frac{n_{ij}}{d_i-1+d_j-1-n_{ij}}$, where $n_{ij}$ is the number of common
neighbors of node $i$ and $j$, and $d_i$ ($d_j$) is the degree of node $i$
($j$)~\cite{Onnela:pnas07}. $O_{ij}$ quantifies the overlap between the
neighbors of two end-points and measures \emph{embeddedness} of an edge.
For instance, $O_{ij} = 0$ indicates that nodes
$i$ and $j$ have no common neighbors and the link is likely to connect
communities. For a large portion of links, overlap decreases with weights. For
a small portion of strongest links ($20\%$ links with $w_{ij}>0.640$ for
Net-sci, $4.3\%$ links with $w_{ij}>3.327$ for Hep-th, $7.8\%$ links with
$w_{ij}>0.765$ for Astro-ph, and $14\%$ links with $w_{ij}>0.869$ for
Cond-mat), overlap increases with weights.  These results indicate that weak
ties mainly constitute dense local clusters, whereas strong ties are connecting
these clusters.
In order to further confirm the universality of weight-topology coupling
patterns in scientific collaboration networks, we examine
network connectivity under link removal~\cite{Onnela:pnas07, Cheng:jsm10,
Pan:EPL12:ca}. We remove links based on descending or ascending order of link
weights and track the relative size of Largest Connected Component (LCC)
$R_{LCC}$ as a function of the fraction of removed links. Figure
\ref{fig:link_rmv} shows that removing strong links breaks the networks into
disconnected components faster than removing weak links, indicating that strong
links are more important in maintaining global network connectivity. Strong
links connect clusters (Fig. \ref{fig:net-sci}\emph{B}), while weak links
reside inside communities (Fig. \ref{fig:net-sci}\emph{C}).

\section{Model}

Many models have been proposed to explain the known properties of scientific
collaboration networks.  However, these models either do not consider link
weights or do not capture the role of strong ties in maintaining global
connectivity. Some models focus on assortative
mixing~\cite{Newman:PRE03:social, Catanzaro:PRE04:assort}. Some study the
self-organizing evolution of collaboration networks as preferential attachment
and ``rich-get-richer"~\cite{Ramasco:pre04:ca, Barabasi:PA02:sci-net,
Borner:pnas04:ca}. Others emphasize the evolution of
disciplines~\cite{Guimera:science05:team} or social interaction of
scientists~\cite{Sun:scirep13}. The weak-tie hypothesis has often been
considered as an evident truth about networks, and most models that produce
community structures assume so~\cite{Jo:plosone11, Kumpula:PRL07:comm}.

By contrast, our model is based on the following observations:
(i) scientific collaboration networks grow in time, as new papers and
scientists join continuously; (ii) junior scientists become inactive with high
probability; indeed, recent work on analysis of the APS dataset reveals that
$40\%$ of authors only publish one
paper in their entire career~\cite{Pan:EPL12:ca}; and (iii) long-term collaboration
usually occurs between senior scientists who have their own research
groups~\cite{Pan:EPL12:ca}.

Our model has two mechanisms of producing new papers: intra-group and inter-group
collaboration. Starting with a research group of an advisor and a student, the
collaboration network grows over time. At every time step: 

\begin{itemize}

\item With probability $c$, each group publishes one paper by itself. The parameter
$c$ controls the ratio of total number of authors to total number of papers. Each paper is written by the
advisor and $l-1$ co-authors preferentially chosen from the
same group based on the students' scientific expertise $e$. The probability to
be chosen is proportional to $e$. If a student joins a group at time $\tau$,
with initial expertise $e(\tau) = 1$, $e$ increases linearly with time: $e(t) =
t - \tau + 1$;

\item Each group may publish up to 
$\alpha$ papers with another
group.  If the group has not had any external collaboration, it chooses
a group randomly and establish a permanent preferred collaboration
relationship. The group tries to write 
$\alpha$ papers, each with probability $c$; Each paper still have $l$ authors, among which two are the
two advisors from each group and $l-2$ are randomly chosen from the pool of
students of the two groups with probability proportional to their expertise;
The parameter $\alpha$ controls the ratio of inter-group collaborations to that
of intra-group. 

\item Each group has one new student;

\item When the expertise reaches a threshold $G$, the student forms a new group
with probability $f$ or becomes inactive with probability $1-f$.

\end{itemize}

\subsection{Simulation Results}

In setting the parameter values (or distributions), we incorporate as many
empirically measured values as possible. First, we
assume the following parameters to be constants, namely $c=0.4$, $G=7$,
and $f=0.2$ in our
analysis. The choice for $c=0.4$ is based on the ratio of total number of
scientists to that of papers in the APS dataset. The other parameters $G$ and
$f$ are also chosen based on real-world observations~\cite{note:para}.
The number of authors in each paper, $l$, is a random variable with the
underlying probability distribution obtained from the APS dataset. The only
free parameter is $\alpha$ and we have performed a robustness analysis of
$\alpha$ and the other parameters in Section~\ref{subsec:robust}.
Table~\ref{tab:parameters} shows the meanings of the model parameters.

\begin{table}
\centering
\begin{tabular}{|c|c|}
\hline
Parameter & Meaning                                            \\ \hline
$c$ & Probability to publish paper                 \\ \hline
$l$       & Number of authors in each paper                    \\ \hline
$G$       & Expertise threshold for students to graduate       \\ \hline
$f$       & Probability of graduates to form new groups        \\ \hline
$\alpha$  & Ratio of inter-group to intra-group collaborations \\ \hline
\end{tabular}
\caption{Model parameters and their explanations.}
\label{tab:parameters}
\end{table}

\begin{figure}[t]
\begin{center}
\includegraphics[trim=0mm 5mm 0mm 0mm, width=\columnwidth]{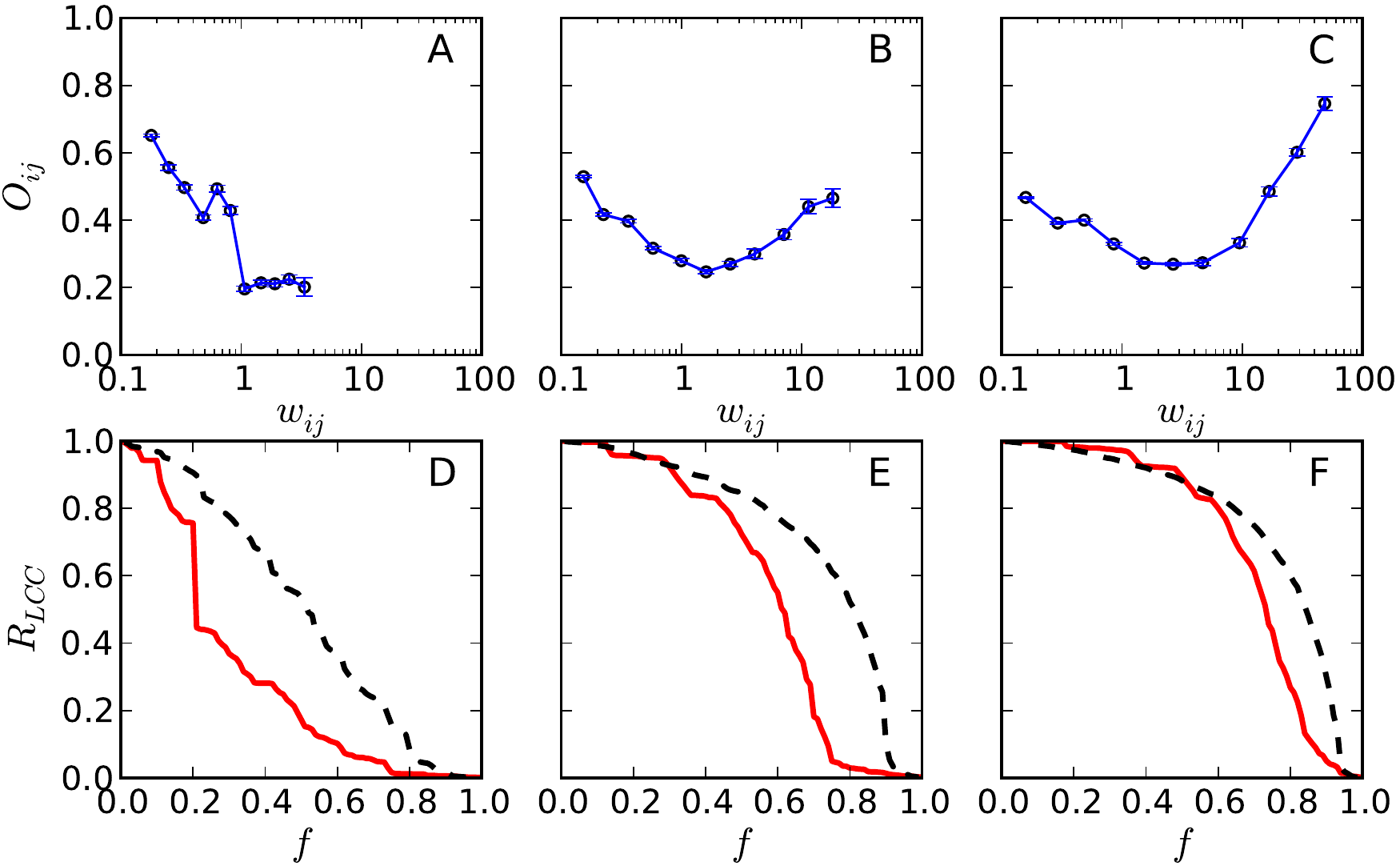}

\caption{\label{fig:mdl-results-corre-robust}Model results with different $\alpha$. Top: correlation between $O_{ij}$ and $w_{ij}$; Bottom: model network robustness to link removal. Left: $\alpha=0$; Middle: $\alpha=1$; Right: $\alpha=3$.}

\end{center}
\end{figure}

We have one free parameter, $\alpha$, and we investigate the impact of
inter-group collaboration by varying it.
Fig.~\ref{fig:mdl-results-corre-robust}\emph{A}-\emph{C} demonstrate that the
v-shaped pattern between overlap and weight can be reproduced when $\alpha \ge
1$, \emph{i.e.}, when there are active inter-group collaborations.
Fig.~\ref{fig:mdl-results-corre-robust}\emph{D}-\emph{F} show that, on the
other hand, the strong ties increasingly maintain global connectivity if we
decrease $\alpha$. As we will demostrate in the next section, when $\alpha \simeq
3.44$, the difference between intra- and inter-group tie strength is 0. Our
  model seems to exhibit the most similar weight organization with the real
collaboration networks around $\alpha = 1$. All the results below are obtained
when $\alpha = 1$. These results indicate that the inter-group collaborations
plays an important role in explaining the atypical tie-strength distributions
in scientific collaboration networks.

\begin{figure}[t]
\begin{center}
\includegraphics[width=\columnwidth]{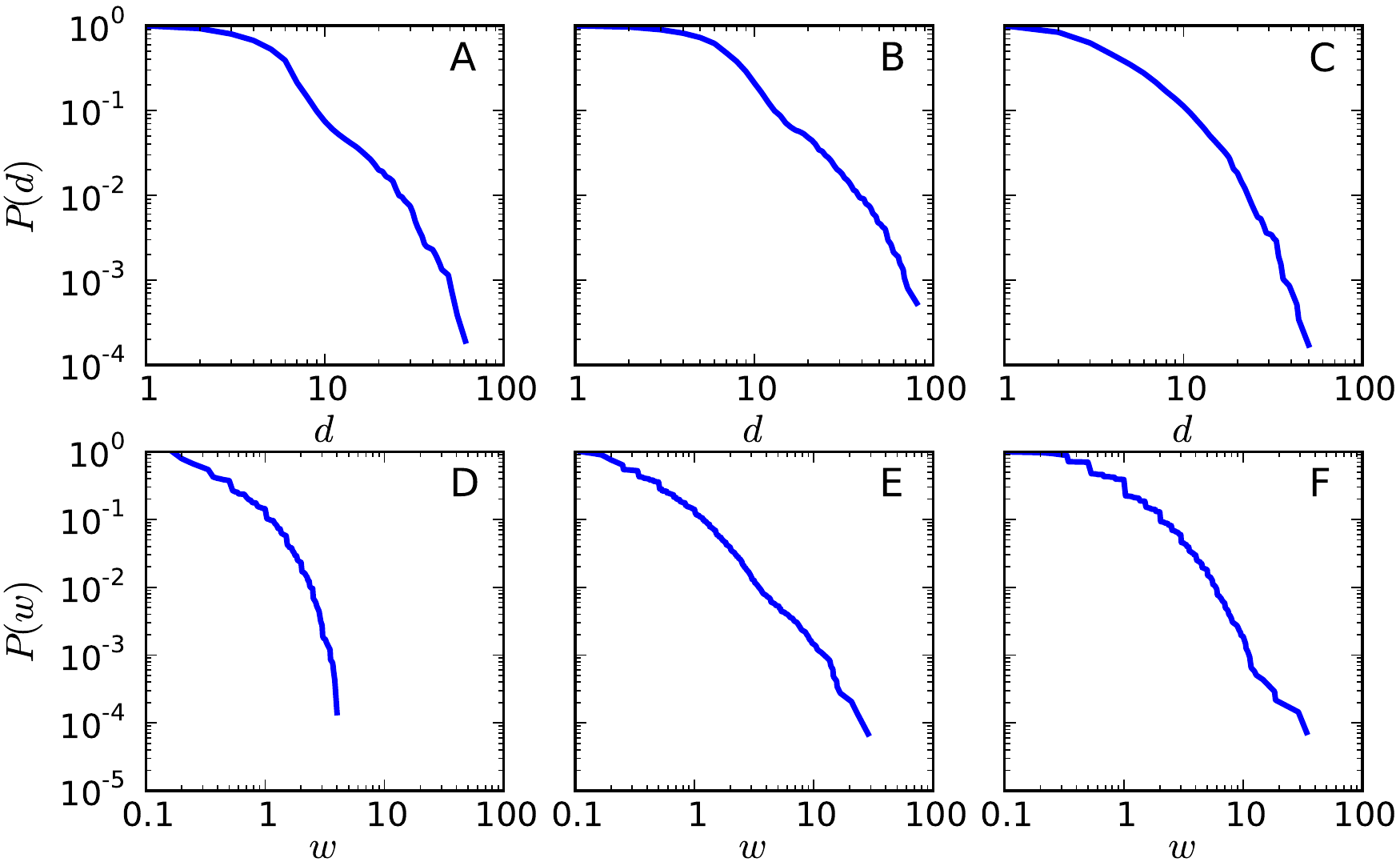}

\caption{\label{fig:mdl-results-deg-wei-dist}Our model produces skewed
degree and weight distributions. Complementary cumulative (Top) degree and
(Bottom) weight distributions. Left: $\alpha=0$; Middle: $\alpha=1$; Right:
Hep-th.}

\end{center}
\end{figure}

Furthermore, our model reproduces other common properties of scientific
collaboration networks: (i) skewed distribution of degree and link weights, as
shown in Fig.~\ref{fig:mdl-results-deg-wei-dist} and (ii) strong clustering
(average clustering coefficient is $0.55$ when $\alpha=1$).

\subsection{Analytical Results}

By calculating the
gained tie strength within a group and between groups at each time step, we show
that when $\alpha \simeq 3.44$, the difference between intra- and inter-group tie strength is 0.
We focus on stationary groups with $G$
students and with total expertise $G(G+1)/2$. With probability $c$, the advisor $a$ in group $g$ writes one paper with the group members. It
will add the weight of $\frac{1}{l-1}$ to the link between the advisor and a
chosen student. Let $p_i(e)$ be the probability that a student with expertise
$e$ is chosen in an intra-group paper (see Appendix~\ref{appendix:prob} for
its calculation). Then the gained link weight between the advisor and the
student is
\begin{equation} w_{a,e}^{(i)} =  \frac{c}{l-1} p_i(e). \end{equation}
Let $p_i(e_1,e_2)$ be probability that two students in the group $g$ with
expertise $e_1$ and $e_2$ are chosen in an intra-group paper. Then the gained 
link weight between the two students is
\begin{equation} w_{e_1,e_2}^{(i)} =  \frac{c}{l-1} p_i(e_1,e_2).
\end{equation}
Meanwhile, the advisor $a$ in group $g$ writes on average $\alpha c$ papers with another 
group $g'$. The weight between the two advisors increases by
\begin{equation} w_{a,a'} = \frac{\alpha c}{l-1}. \end{equation}
Let $p_b(e)$ be the probability that a student with expertise $e$ is chosen in an inter-group paper. Then the expected gain, through inter-group collaborations, of weights between an advisor and a student with expertise $e$ either in the same group ($w_{a,e}^{(i')}$) or the other group ($w_{a,e}^{(b)}$) are represented as follows: 
\begin{equation} w_{a,e}^{(i')} = w_{a,e}^{(b)} = \frac{\alpha c}{l-1} p_b(e). 
\end{equation}
Let $p_b(e_1,e_2)$ be the probability that two students with expertise $e_1$ and
$e_2$ are chosen in an inter-group paper. Then
\begin{equation}
w_{e_1,e_2}^{(i')} = w_{e_1,e_2}^{(b)} = \frac{\alpha c}{l-1} p_b(e_1,e_2).
\end{equation}
So the gained tie strength within a group at each time step is
\begin{equation}
W^{(i)} = \sum_{e=1}^G w_{a,e}^{(i)} + \sum_{e_1 \neq e_2} w_{e_1,e_2}^{(i)} + \sum_{e=1}^G w_{a,e}^{(i')} + \sum_{e_1 \neq e_2} w_{e_1,e_2}^{(i')}.
\end{equation}
The total gained tie strength between the two groups is
\begin{equation}
W^{(b)} = w_{a,a'} + \sum_{e=1}^G w_{a,e}^{(b)} + \sum_{e_1=1}^G \sum_{e_2=1}^G w_{e_1,e_2}^{(b)}.
\end{equation}
The difference between inter-group tie strength and intra-group tie strength is
\begin{widetext}
\begin{eqnarray}
\label{eq:delta_w}
\Delta W &=& W^{(i)} - W^{(b)} \nonumber \\
&=& \frac{c}{l-1} \left( \sum_{e=1}^G p_i(e) + \sum_{e_1 \neq e_2} p_i(e_1,e_2) - \alpha - \alpha \sum_{e=1}^G p_b(e,e) \right)
\end{eqnarray}
\end{widetext}

$\Delta W = 0$ when
\begin{equation}
\alpha_c = \frac{\sum_{e=1}^G p_i(e) + \sum_{e_1 \neq e_2} p_i(e_1,e_2)}{1 + \sum_{e=1}^G p_b(e,e)} \simeq 3.44.
\end{equation}
Indeed, Fig.~\ref{fig:mdl-results-corre-robust}\emph{F} shows that with $\alpha = 3$,
the removal of weak and strong ties similarly affects the connectivity of the network until about 60\% of the edges removed.

\begin{figure}[t]
\begin{center}
\includegraphics[trim=5mm 6mm 0mm 5mm, width=\columnwidth]
{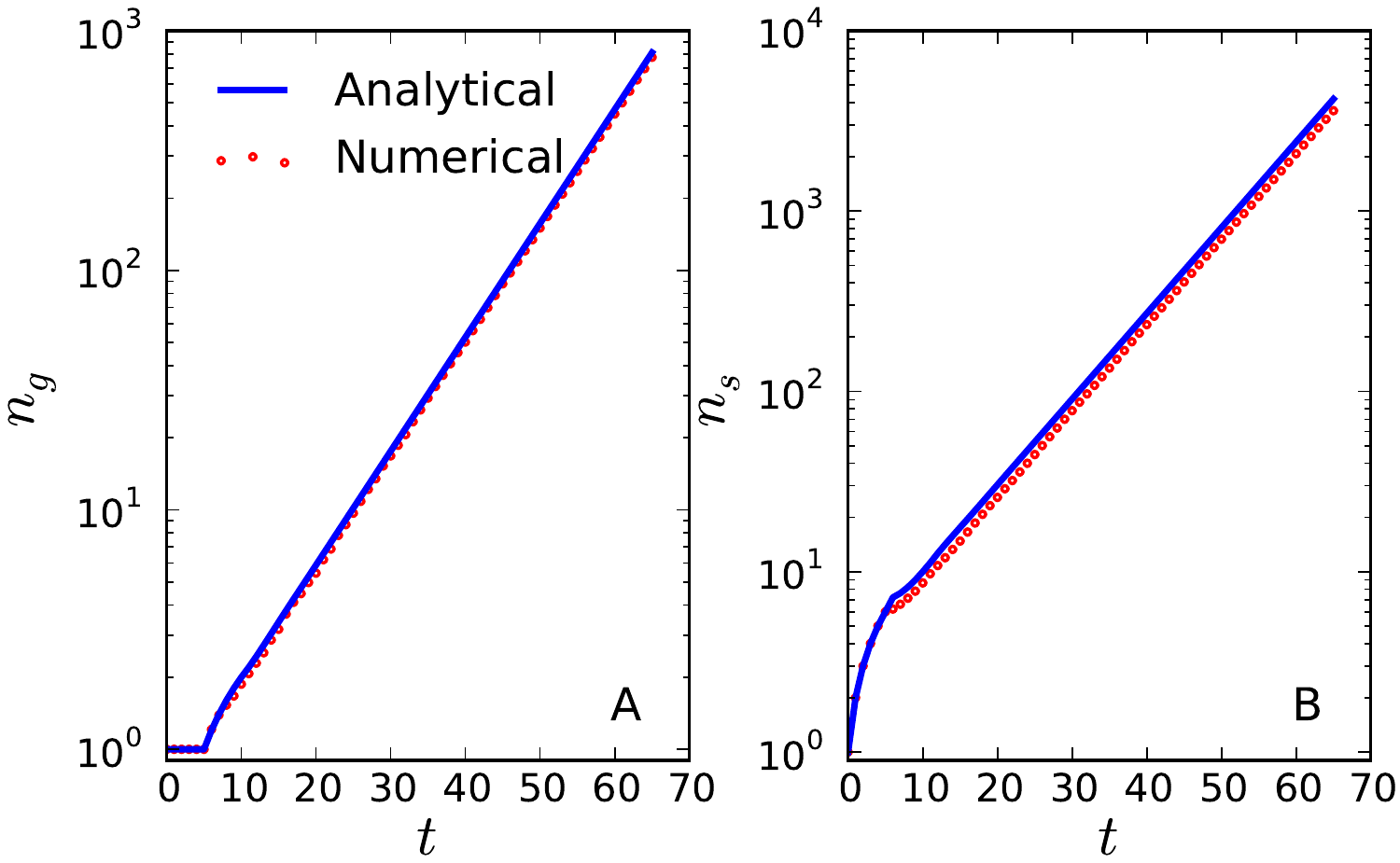}

\caption{\label{fig:comparison}Comparison
of calculation results with numerical results of (\emph{A}) number of of groups
$n_g(t)$ and (\emph{B}) number of students $n_s(t)$.  The numerical results are
averaged over $100$ repetitions.}

\end{center}
\end{figure}

We next derive the number of groups $n_g(t)$ and the number of students $n_s(t)$
at $t$. Let $n_s^{(e)}(t)$ be the number of students at the expertise level $e$
at time step $t$. The expertise $e$ increases in time and the students graduate
when the expertise reaches $G$. Graduates create their own group with
the probability $f$, namely
\begin{equation} n_g(t) =  n_g(t-1) + f n_s^{(G-1)}(t-1), \end{equation}
with $n_g(0) = \ldots = n_g(G-2) = 1$. At each time step, there are the same
number of new students as the number of groups
\begin{equation} n_s^{(1)}(t) = n_g(t). \end{equation}
The number of students with expertise $e \geq 2$ is the same as the number of
students with expertise $e-1$ in the previous time step
\begin{equation} n_s^{(e)}(t) = \ldots = n_s^{(1)}(t-(e-1))
= n_g(t-e+1). \end{equation}
Therefore, the number of groups is
\begin{align} n_g(t) &= n_g(t-1) + fn_g(t-G+1).\ &(t \geq G-1) \end{align}
The number of students is
\begin{align} n_s(t) &= \sum_{e=1}^G n_s^{(e)}(t) = \sum_{e=1}^G n_g(t-e+1).\
&(t \geq G-1) \end{align}
with $n_s(t) = t+1$ for $t \leq G-2$. The increased number of nodes is the
number of groups in the previous time step
\begin{equation} N(t) = N(t-1) + n_g(t-1) \end{equation}
with $N(0)=0$ and $N(t)=t+1$ for $1 \leq t \leq G-2$. These analytical results are
in agreement with the numerical results, as shown in Fig. \ref{fig:comparison}.

Finally, we offer the calculation of the mean degree $\left< d \right>$ of model networks
in Appendix~\ref{appendix:deg}.

\subsection{Robustness Analysis} \label{subsec:robust}

\begin{figure}[t]
\begin{center}
\includegraphics[trim=0mm 5mm 0mm 0mm, width=\columnwidth]{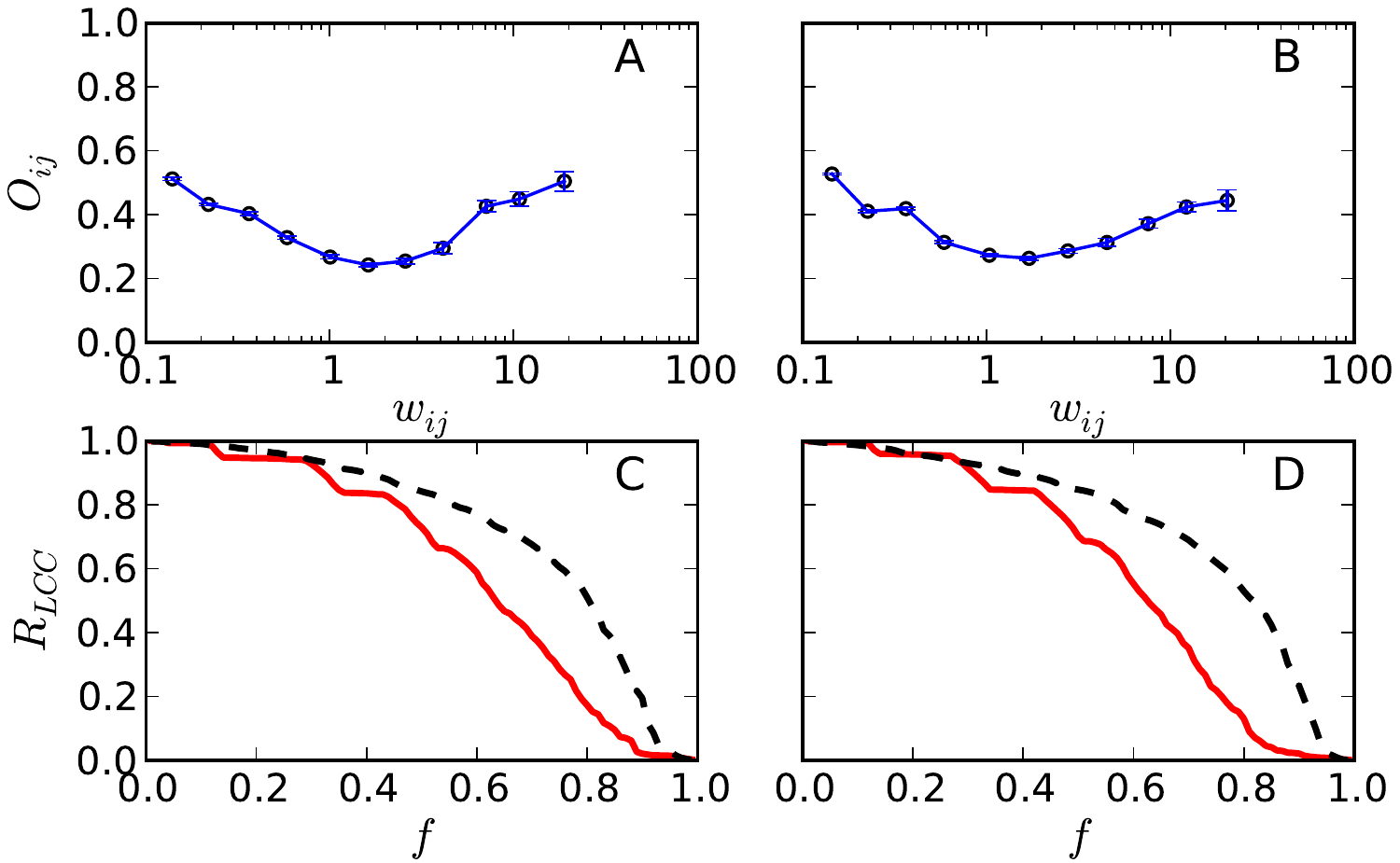}

\caption{\label{fig:mdl-robust-g}(Top) Weight-overlap correlation and (Bottom) robustness to link removal when (Left) $G = 6$ and (Right) $G = 8$.}

\end{center}
\end{figure}

\begin{figure}[t]
\begin{center}
\includegraphics[trim=0mm 5mm 0mm 0mm, width=\columnwidth]{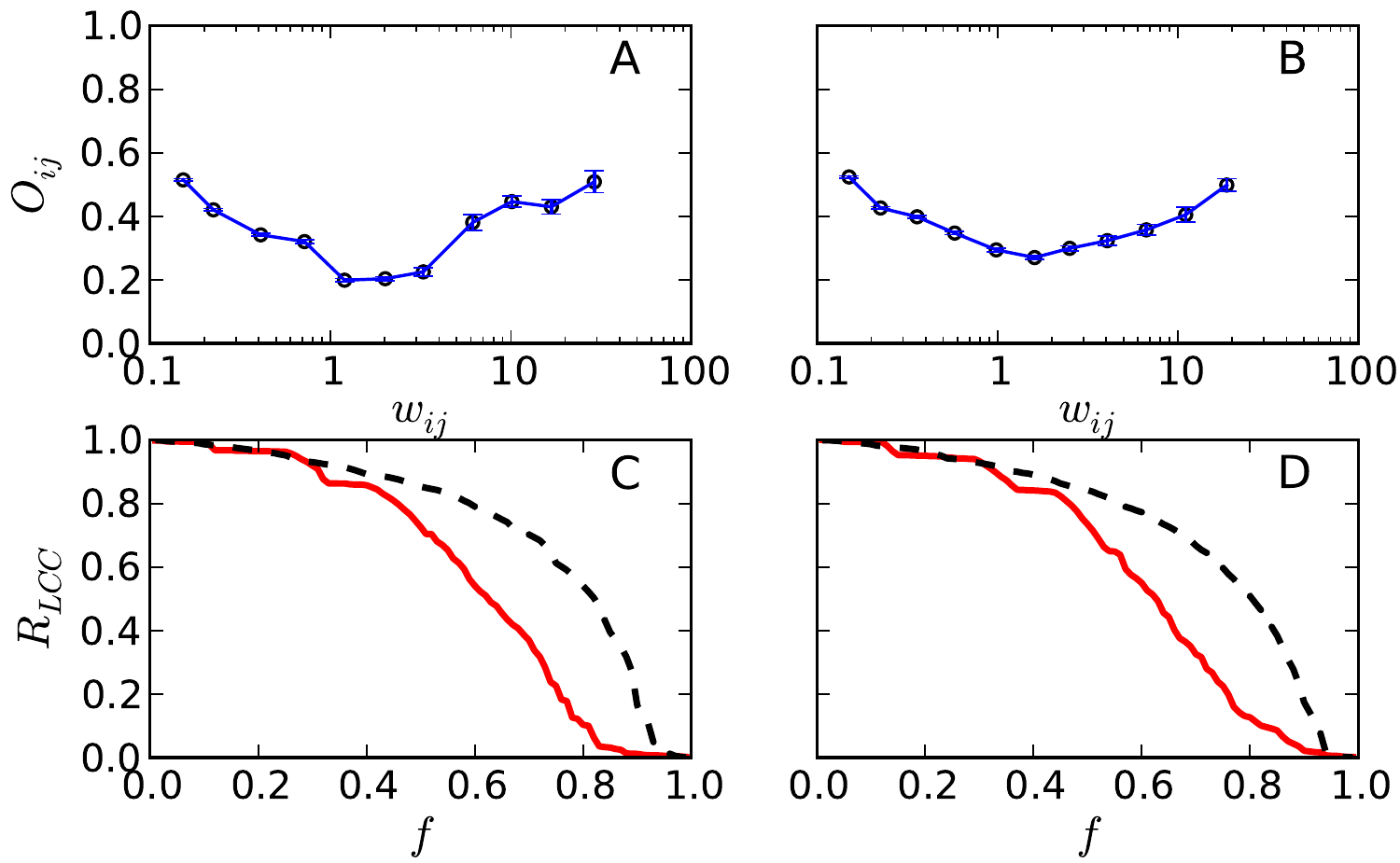}

\caption{\label{fig:mdl-robust-f}(Top) Weight-overlap correlation and (Bottom) robustness to link removal when (Left) $f = 0.1$ and (Right) $f = 0.3$.}

\end{center}
\end{figure}

We now investigate the model's sensitivity to the parameters $G$ and $f$. Fig.~\ref{fig:mdl-robust-g} and~\ref{fig:mdl-robust-f} show that our model is robust to the choices of parameters $G$ and $f$. The two observations are still produced when (Fig.~\ref{fig:mdl-robust-g}) $G = 6$ or $G = 8$ and when (Fig.~\ref{fig:mdl-robust-f}) $f = 0.1$ or $f = 0.3$.

\section{Conclusions}

In this paper we explore the weight organization of scientific collaboration
networks.  We propose a model, which incorporates intra- and inter-group
collaborations and reproduces many properties of real-world collaboration
networks.  We also provide detailed analysis of our model. Our work also raises
further questions such as: How did the collaboration pattern change in time?
How do scientific ideas flow through strong and weak ties? Are there any
general coupling patterns (or classes) between structure and
weights?

\begin{acknowledgments}

The authors would like to thank Lilian Weng, Nicola Perra, M\'arton Karsai, Filippo Menczer,
Alessandro Flammini, Filippo Radicchi, Martin Rosvall, Jie Tang, and Honglei Zhuang for helpful discussions,
and John McCurley for editorial assistance.

\end{acknowledgments}

\appendix

\section{Calculation of mean degree} \label{appendix:deg}
In order to get the expected degree of an advisor when graduation, we track it from time step
$\tau$ when she joined a stationary group $g$ as a student $a_1$ with initial expertise $e=1$
to $\tau+G-1$ when graduation with expertise $G$. For intra-group collaboration, there are
already a total of $G-2$ students in group $g$ at $\tau$. From $t=\tau+1$ to $t=\tau+G-2$,
there are another $G-2$ new students joining in the group. Let $I(a_1, a_2)$
($\text{Pr}(a_1,a_2)$) be the indicator function (probability) that $a_1$ have collaborated with
$a_2$. The number of different students collaborators in the group $g$ for $a_1$ is
\begin{eqnarray}
\label{eq:intra-deg}
d_i &=& \sum_{a_e \in g} I(a_1, a_e) = \sum_{a_e \in g} 1 \times
\text{Pr}(a_1, a_e) \nonumber \\
&=& \sum_{a_e \in g} 1 - \overline{\text{Pr}}(a_1, a_e) \nonumber \\
&=& 2(G-2) - \sum_{a_e \in g} \overline{\text{Pr}}(a_1, a_e).
\end{eqnarray}

Consider the student $a_2$ with expertise $e=2$ at time $t=\tau$, $a_1$ and $a_2$ will be in
the same group until the expertise of $a_2$ reaches $G-1$. Let $\bar{p}(e_1, e_2)$ be
the probability that two students with expertise $e_1$ and $e_2$ are not chosen in the one intra-group paper
\begin{equation} \bar{p}(e_1, e_2) = (1-c) + c(1-p_i(e_1,e_2)). \end{equation}
Then $\overline{\text{Pr}}(a_1,a_2)$ is the probability that $a_1$ do not collaborate with
$a_2$ from $t=\tau$ to $t=\tau+G-3$
$$\overline{\text{Pr}}(a_1,a_2) = \prod_{e=1}^{G-2} \bar{p}(e,e+1).$$
Similarly, for student $a_3,\ldots,a_{G-1}$ with expertise $e=3,\ldots,G-1$ at time $t=\tau$
\begin{eqnarray}
\overline{\text{Pr}}(a_1,a_3) &=& \prod_{e=1}^{G-3} \bar{p}(e,e+2), \nonumber \\
&\ldots& \nonumber \\
\overline{\text{Pr}}(a,a_{G-1}) &=& \prod_{e=1}^{1} \bar{p}(e,e+G-2). \nonumber
\end{eqnarray}

From time step $t=\tau+1$ to $t=\tau+G-2$, the expertise of $a_1$ increases from $e=2$
to $G-1$. A new student joins the group at each time step
\begin{eqnarray}
\overline{\text{Pr}}(a_1,a_G) &=& \prod_{e=2}^{G-1} \bar{p}(e,e-1), \nonumber \\
&\ldots& \nonumber \\
\overline{\text{Pr}}(a_1,a_{2G-3}) &=& \prod_{e=G-1}^{G-1} \bar{p}(e,e-(G-2)). \nonumber
\end{eqnarray}

The intra-group degree for student collaborators (Eq.~\ref{eq:intra-deg}) now is
\begin{widetext}
\begin{equation}
\label{eq:intro-deg-result}
d_i = 2(G-2) - \sum_{j=1}^{G-2} \prod_{e=1}^{G-1-j} \overline{p}(e,e+j) -
\sum_{j=2}^{G-1} \prod_{e=j}^{G-1} \overline{p}(e,e-(j-1)).
\end{equation}
\end{widetext}

The inter-group degree $d_b$ is similar except different probability form. There
are a total of $2(G-2)+1$ number of different students (one more student in group
$g'$ with expertise $1$). Let $\bar{p}_b^{(\alpha)}(e_1,e_2)$ be the probability
that two students with expertise $e_1$ and $e_2$ are not chosen in any of the
$\alpha$ inter-group papers. Then, the number of student collaborators in another group $g'$ is
\begin{widetext}
\begin{equation}
\label{eq:inter-deg-result}
d_b 
= 2G-3 - \sum_{j=0}^{G-2} \prod_{e=1}^{G-1-j} \overline{p}^{(\alpha)}(e,e+j) - 
\sum_{j=2}^{G-1} \prod_{e=j}^{G-1} \overline{p}^{(\alpha)}(e,e-(j-1)).
\end{equation}
\end{widetext}

Considering two advisers, we get the degree of the student $a_1$ when graduation
\begin{equation} d = d_i + d_b + 2. \end{equation}

After graduation, $a_1$ becomes an advisor with
probability $f$. The increased degree at each time step now is the probability
of collaborating with the two newly joined students
\begin{equation} \Delta d_m = \left( 1-(1-p_i(1)) \right) +
\left(1-(1-p_b(1))^{\alpha} \right).  \end{equation}

The total degree of advisors at time step $t$ is
\begin{equation}
D_m(t) = \sum_{\tau=G-1}^{t-1} \left\{n_s^{(G-1)}(\tau) \cdot f \cdot
\left[d+\left(t-\tau+1\right)\Delta d_m\right] \right\}.
\end{equation}

Using the same idea, we can get the degree of a student with expertise $e$ ($e \leq G-1$) at
time step $t$. This is a general case of Equations~\ref{eq:intro-deg-result}
and~\ref{eq:inter-deg-result}
\begin{widetext}
$$d_i^{(e)}(t) = 2(G-2) - \sum_{i=1}^{G-1-e} \prod_{j=1}^{e} \bar{p}(j,j+i) -
\sum_{i=1}^{e-1} \prod_{j=1}^{e-i} \bar{p}(j,j+G-1-(e-i)) - 
\sum_{i=2}^{e} \prod_{j=i}^{e} \bar{p}(j,j-(i-1)),$$
$$d_b^{(e)}(t) = 2G-3 - \sum_{i=0}^{G-1-e} \prod_{j=1}^{e} \bar{p}^{(\alpha)}(j,j+i) - 
\sum_{i=1}^{e-1} \prod_{j=1}^{e-i} \bar{p}^{(\alpha)}(j,j+G-1-(e-i)) - 
\sum_{i=2}^{e} \prod_{j=i}^{e} \bar{p}^{(\alpha)}(j,j-(i-1)).$$
\end{widetext}
\begin{equation} d^{(e)}(t) = d_i^{(e)}(t) + d_b^{(e)}(t) + 2. \end{equation}

So the mean degree $\left< d \right>$ at time step $t$ is
\begin{equation}
\left< d \right>(t) = \frac{1}{N(t)} [D_m(t) + \sum_{e=1}^{G-1} n_s^{(e)}(t) d^{(e)}(t)].
\end{equation}

\section{Calculation of $p_i(e)$} \label{appendix:prob}
We describe how to calculate $p_i(e)$, $p_b(e)$, $p_i(e_1,e_2)$, and $p_b(e_1,e_2)$. 
Recall that $p_i(e)$ is the probability that a student with expertise $e$ is chosen for one
intra-group paper. Its distribution is the sum of multiple multivariate Wallenius' noncentral
hypergeometric distributions~\cite{Chesson:hypergeometric} and can be calculated by using
package \texttt{BiasedUrn} in R~\cite{BiasedUrn}. The calculations for $p_b(e)$,
$p_i(e_1,e_2)$, and $p_b(e_1,e_2)$ are similar.

\bibliographystyle{apsrev}
\bibliography{ca-tie}

\begin{thebibliography}{46}
\bibitem[{not()}]{note:para}
\bibinfo{note}{The expertise threshold for students to graduate, $G$, reflects
  the median year to earn a Ph.D. degree reported in~\cite{nsf:year};
  probability of graduates to form new groups, $f$, is based on the percentage
  of Ph.D. students who land academic jobs by graduation~\cite{atlantic:job}.}
\end{thebibliography}

\end{document}